\documentclass[]{aa}
\usepackage[]{psfig}
\usepackage{txfonts}
\begin{document}

\title{A photometric monitoring of bright high-amplitude $\delta$ Scuti stars}
\subtitle{II. Period updates for seven stars}

\author{A. Derekas\inst{1,2,5} \and L. L. Kiss\inst{2,5}\thanks{On leave from
University of Szeged, Hungary} \and P. Sz\'ekely\inst{1}
\and E. J. Alfaro\inst{3}
\and B. Cs\'ak\inst{1} \and Sz. M\'esz\'aros\inst{1} \and 
E. Rodr\' \i guez\inst{3}
\and A. Rolland\inst{3} \and K. S\'arneczky\inst{4,5} \and 
Gy. M. Szab\'o\inst{1} \and K. Szatm\'ary\inst{1} \and M. V\'aradi\inst{1}
\and Cs. Kiss\inst{6}} 

\institute{Department of Experimental Physics and Astronomical Observatory,
University of Szeged, Szeged, D\'om t\'er 9., H-6720 Hungary
\and School of Physics, University of Sydney 2006, Australia
\and Instituto de Astrof\'\i sica de Andaluc\'\i a, CSIC, P.O. Box 3004,
18080-Granada, Spain
\and Astronomical Observatory, Szeged, Hungary
\and Guest observer at Konkoly Observatory
\and Konkoly Observatory of the Hungarian Academy of Sciences, P.O. Box 67,
H-1525 Budapest, Hungary
}

\titlerunning{New observations and period updates for 7 HADS}
\authorrunning{A. Derekas et al.}
\offprints{L. L. Kiss,\\
e-mail: {\tt laszlo@physics.usyd.edu.au}}
\date{}

\abstract{We present new photometric data for seven high-amplitude $\delta$ Scuti
stars. The observations were acquired between 1996 and 2002, mostly in  
the Johnson photometric system. For one star (GW~UMa), our observations are the 
first since the discovery of its pulsational nature from the 
Hipparcos data.The primary goal of this
project was to update our knowledge on the period variations of the target stars.
For this, we have collected all available photometric observations
from the literature and constructed decades-long O$-$C diagrams of the
stars. This traditional method is useful because of the single-periodic 
nature of the light variations. Text-book examples of slow period evolution 
(XX~Cyg, DY~Her, DY~Peg) and cyclic period changes due to light-time effect (LITE) 
in a binary system (SZ~Lyn) are updated with the new
observations. For YZ~Boo, we find a period decrease instead of increase.
The previously suggested LITE-solution of BE~Lyn (Kiss \& Szatm\'ary 1995)
is not supported with the new O$-$C diagram. Instead of that, we suspect the 
presence of transient light curve shape variations mimicking small period 
changes.
} 
 
\maketitle

\section{Introduction}

High-amplitude $\delta$ Scuti stars (hereafter HADSs) are either Pop. I
stars close to the main sequence or evolved Pop. II stars (these are also
known as SX~Phe variables) with very characteristic light variations
caused by radial pulsations (Rodr\'\i guez et al. 1996). Typical 
periods range from 0\fd05 to 0\fd15 associated with relatively 
large amplitudes ($A_{\rm V}\geq0\fm30$ is a widely adopted convention). Owing 
to their short periods and high amplitudes, these stars are very good 
targets for small and moderate-sized telescopes, so that interesting
astrophysical phenomena can be easily studied even with modest instrumentation.
Apart from attempts to detect evolutionary effects, the most 
interesting case studies are related to the suspected binary HADSs, in which
the physical parameters of the pulsating component can be constrained from
the binary nature. However, the time-span of the available observations is often
too short, therefore, significant observational efforts have to be done to 
reach unambiguous conclusions.  

In Paper I (Kiss et al. 2002b) we discussed the 
double-mode pulsation of V567~Ophiuchi. The aims of this project have 
been outlined in Kiss et al. (2002b) and to avoid repetition, we 
refer to theoretical and observational aspects discussed 
in that paper. The main aim of this paper is to publish a new 
dataset for seven variable stars. Besides the new photometric data 
we give an updated description of their period changing behaviour. 
The analysis of multicolour data in terms of physical parameters 
will be published in the final paper of the series (Derekas et al., 
in prep.).

The structure of the paper is the following. Target selection and observations
are described in Sect.\ 2. The main part of the paper is Sect.\ 3, in which the
results are presented. Separate subsections are devoted to individual stars,
where every relevant detail (observational data, sample light curves,
updated O$-$C diagrams) is given. A brief summary is presented in Sect.\ 4.

\section{Target selection and instrumentation}

The observations were started with regular monitoring of BE~Lyncis, for which we
suspected binarity (Kiss \& Szatm\'ary 1995). The first extension was made toward
XX~Cygni (Kiss \& Derekas 2000) and since then, we have tried to include  all
HADSs in the northern sky brighter than $V=11\fm3$ at maximum (this limit was
determined by the typical accuracy and limiting magnitudes of our main
instruments). Unfortunately, this goal was unreachable because of unfavourable
weather conditions  in several observing seasons. That is why we excluded the
following bright northern HADSs: GP~And, AD~CMi, DH~Peg, V1719~Cyg. Also, we
concentrated on the monoperiodic variables and some of these excluded stars are
well-known double-mode pulsators (see Rodr\'\i guez et al. 1996). Thus, our
sample consisted of 10 variables, however, two stars (V1162~Ori and DE~Lac) had
so meagre coverage that we had to remove them from the final target list.
V567~Oph turned to be a  double-mode pulsator and was discussed in Kiss et al.
(2002b). The remaining seven HADSs are listed in Table\ 1, where their main
observational properties are also summarized.

\begin{table}
\begin{center}
\caption{The list of programme stars. The magnitude ranges and periods are 
taken from the GCVS.}
\begin{tabular}{|llrrlc|}
\hline
Star & Pop. & $V_{\rm max}$ & $V_{\rm min}$ & P (d) & Type of obs.\\
\hline
DY~Her & I  & 10\fm15 & 10\fm66 & 0.14863 & $V$ \\
YZ~Boo & I  & 10\fm30 & 10\fm80 & 0.10409 & $V$ \\
XX~Cyg & II & 11\fm28 & 12\fm13 & 0.13486 & $V(RI)_C$\\
SZ~Lyn & I  & 9\fm08 & 9\fm72 & 0.12053 & $V$\\
BE~Lyn & I  & 8\fm60 & 9\fm00 & 0.09587 & $BV/uvby$\\
DY~Peg & II & 9\fm95 & 10\fm62 & 0.07293 & $BV/uvby$\\
GW~UMa & II? & 9\fm48 & 9\fm97 & 0.20319$^a$ & $BVI_C$\\
\hline
\end{tabular}
\end{center}
$^a$ ESA (1997) 
\end{table}

During the seven years of observations, we have utilized various instrumentations 
at three observatories (Szeged Observatory, Piszk\'estet\H{o} Station of the 
Konkoly Observatory, Sierra Nevada Observatory). In the following we briefly
introduce the telescopes and detectors used in this project. As the primary 
aim was to obtain good light curve coverage enabling accurate 
determination of the epochs of maximum, the CCD measurements were  
uninterrupted single-filtered $V$-band observations, with a few exceptions. 

\begin{itemize}

\item {\em Szeged Observatory, 0.28m Schmidt-Cassegrain (Sz28)}

This telescope is located in the very centre of the city of Szeged, thus
suffering from strong light pollution. For our observations on two nights 
in 2002, it was equipped with an SBIG ST--7I CCD camera (765$\times$510 
9$\mu$m pixels giving a field of view (FOV) of 11\farcm5$\times$7\farcm5). This
instrument was the least used in our project. 

\item {\em Szeged Observatory, 0.4m Cassegrain (Sz40)}

The majority of the observations were acquired with the 0.4m Cassegrain-telescope
of the Szeged Observatory. Between 1996 and 2000, photoelectric
photometry was carried out through Johnson $BV$ filters. The detector was an 
SSP--5A photoelectric photometer and we obtained differential photometric 
data using selected comparison stars located near to target stars. 
The photometer was replaced by an SBIG ST--9E CCD camera
in 2001 (512$\times$512 20$\mu$m pixels, FOV=$6^\prime\times6^\prime$) and 
most single-filtered $V$-band observations were made with this instrument. 

\item {\em Piszk\'estet\H{o}, 0.6m Schmidt (P60)}

Fewer observations were done with the 60/90/180cm Schmidt-telescope mounted 
at the Piszk\'estet\H{o} Station of the Konkoly Observatory. The detector 
was a Photometrics AT200 CCD camera (1536$\times$1024 9$\mu$m pixels,
FOV$=28^\prime\times19^\prime$). The observations were done either in $B$ or 
$V$ filters.

\item {\em Sierra Nevada Observatory, 0.9m Ritchey-Chr\'etien (SNO90)}

Occasionally, we acquired simultaneous $uvby$ Str\"omgren photometric 
observations using the 0.9m
telescope of the Sierra Nevada Observatory (Spain) equipped with a 
six-channel Str\"omgren-Crawford spectrograph photometer. 

\end{itemize}

The data were reduced in a standard fashion. For the photoelectric 
observations, we made use of different computer programs written by LLK, 
which were based on methods described in Henden \& Kaitchuk (1982). 
The CCD observations were reduced with standard tasks in IRAF\footnote
{IRAF is distributed by the National Optical Astronomy Observatories,
which are operated by the
Association of Universities for Research in Astronomy, Inc., under  
cooperative agreement with the National Science Foundation.}, including bias
removal and flat-field correction utilizing sky-flat images taken during 
the evening or morning twillight. Differential magnitudes were calculated with 
aperture photometry using two comparison stars of similar brightnesses. We have
not tried PSF photometry because the small FOVs contained too few stars for
a reliable PSF-determination (see Vink\'o et al. 2003 for a related discussion 
using the same instruments). The typical photometric accuracy was 
$\pm0\fm01-0\fm02$, depending on the weather conditions and brightness
difference of the variable and comparison stars. In fields far from
the galactic plane -- e.g. those of BE~Lyn and SZ~Lyn -- we had to choose
somewhat fainter comparisons, typically 2--3 mags dimmer than the target, which
consequently increased the noise level of the data. 

Thanks to the dense sampling of the light curves, new times of maximum light 
were easy to determine from the individual cycles. This was done by fitting
low-order (3--5) polynomials to the light curves around maxima. 
We repeated the fitting procedure by changing the parameters of the fits 
to yield some insights into the uncertainty range of the determined epochs. 
We estimate the typical accuracy as $\pm$0\fd0003 (i.e. $\pm$26 sec), which is
comparable with the usual exposure lengths. 

\section{Results}

The period variations were studied with the classical O$-$C (observed 
{\it minus} calculated) method. The use of
this simple approach is justified by the remarkable stability of the 
HADS light curves, both in amplitude (Rodr\'\i guez 1999) and in period
(Rodr\'\i guez et al. 1995, Breger \& Pamyatnykh 1998, hereafter BP98). 
The blind use of the O$-$C technique has been strongly 
criticised by Lombard and Koen in their series of papers (e.g. Lombard \& Koen
1993, Koen \& Lombard 1995, Koen 1996, Lombard 1998) underlining the importance 
of the evaluation of the model residuals. For this, they developed various 
statistical methods by analysing autocorrelation of the residuals. This is
a crucial point when studying long-period stars (e.g. Mira variables) with 
relatively large stochastic scatter of the period. In case of the
HADSs one does not expect such intrinsic period scatter (as suggested by the
observations) and evolutionary period changes might be constrained with the O$-$C
technique. 

Nevertheless, in every case we have checked the model residuals via
their frequency analysis. This can be used to search for 
secondary periodicity or slow changes of the period hidden by the observational
scatter. The basic idea is the fact that a small 
modulation of the O$-$C diagram appears as an additional periodicity 
of the O$-$C residuals. A recent example for this has been given 
by P\'ocs et al. (2002), who presented the Fourier spectrum of the O$-$C diagram of
the multiply periodic HADS RV~Ari, clearly revealing $f_{\rm m}=f_1-f_0$ in their
data. On the other hand, if there is a slight period change in the data
and a wrong model is subtracted, then the residuals' spectrum will 
show a peak very close to $f={1 \over P}$. This closely separated peak in 
the Fourier spectrum, however, can be explained by another way, too: it
appears simply because of the periodicity of the spectrum itself. The 
Fourier transform of an O$-$C diagram is periodic with the frequency of the
ephemeris because of the periodicity of data sampling. Therefore, if a 
long-term trend is present in the residuals, that causes a low-frequency 
component at $f_{\rm low}$, which appears again at $f_{\rm low}+f$.
Whatever explanation we adopt, Fourier analysis of the residuals can 
help emphasize the hardly visible trends of the data.

\subsection{DY~Herculis}

\begin{figure}
\begin{center}
\leavevmode
\psfig{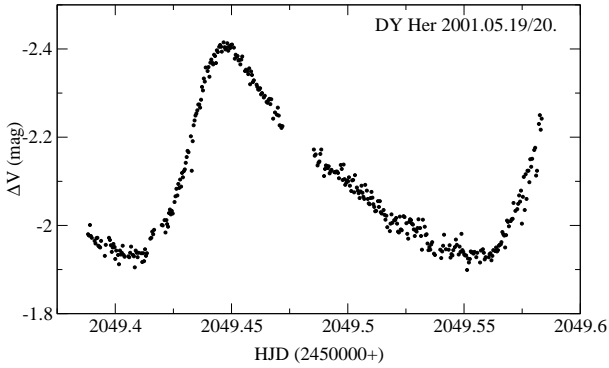}
\caption{Sample light curve for DY~Her.}
\label{dyhlc}
\end{center}
\end{figure}

The most recent period-change study for DY~Her was made by Yang et al. (1993), but 
even the last data of that paper were obtained in 1984. Previously, Szeidl \&
Mahdy (1981) collected all available data to study long-term period
variations. Both studies revealed a slow period decrease of the star. 
The main characteristics are summarized by Pe\~na et al. (1999). 
Recently, P\'ocs \& Szeidl (2000) interpreted the long-term O$-$C diagram
in terms of the light-time effect (LITE) caused by a hypothetical low-mass
component, although their conclusions were quite ambiguous.

We observed the star on four nights in 2001 with Sz40. More than 2000 individual
exposures of 20--40 seconds were obtained in $V$-band. Two comparison stars
located within 6$^\prime$ were used (comp=GSC 0968-1532, check=GSC 0968-1002, both
stars are about 11\fm0). The full log is given in Table\ \ref{dyhobs}, a
sample light curve is shown in Fig.\ \ref{dyhlc}. There is no new time of maximum
for the first night, because we had to interrupt our observations for half an
hour and the maximum just appeared during that break.

\begin{table}
\begin{center}
\caption{Journal of observations and new times of maximum for DY~Her}
\label{dyhobs}
\begin{tabular}{|lllrl|}
\hline
Date       & filter & Inst. & Points & $HJD_{\rm max}$\\
\hline
2001-05-09 & $V$    & Sz40  & 509 & --\\
2001-05-10 & $V$    & Sz40  & 663 & 2452040.3804\\
           &        &       &     & 2452040.5317\\
2001-05-19 & $V$    & Sz40  & 409 & 2452049.4471\\
2001-06-25 & $V$    & Sz40  & 506 & 2452086.4579\\
\hline
\end{tabular}
\end{center}
\end{table}

\begin{figure}
\begin{center}
\leavevmode
\psfig{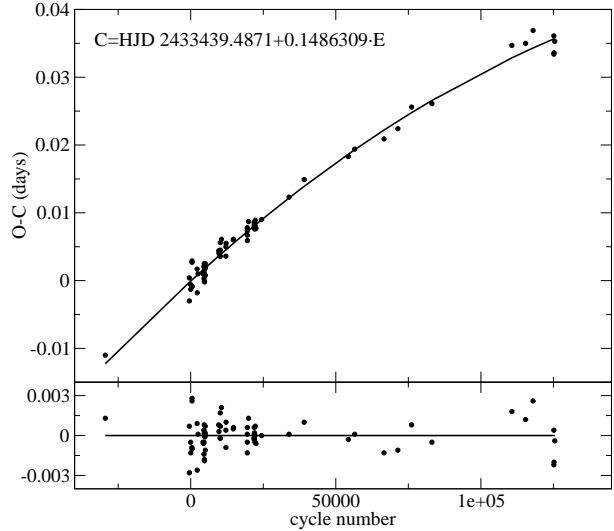}
\caption{The O$-$C diagram of DY~Her. The solid line shows the 
parabolic fit.}
\label{dyhoc}
\end{center}
\end{figure}

We have collected all times of maximum from the literature (the updated list
of references is given in P\'ocs \& Szeidl 2000) to arrive to the final O$-$C
diagram. It is plotted in Fig.\ \ref{dyhoc} and was 
calculated with the following ephemeris:

$$HJD_{\rm max}=2433439.4871+0.1486309 \times E$$

\noindent (the epoch is from the GCVS, the period from ESA 1997).

\begin{figure}
\begin{center}
\leavevmode
\psfig{figure=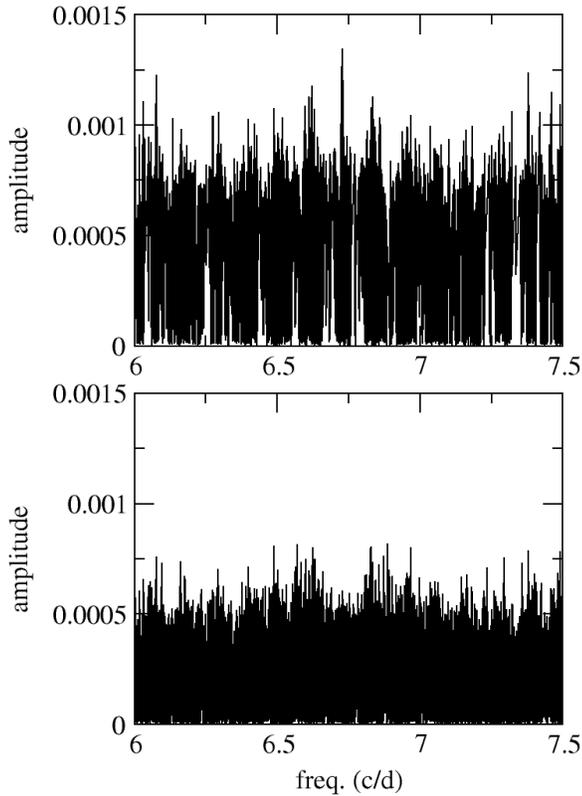,width=8cm}
\caption{Top panel: the Fourier spectrum of the residuals for DY~Her
after subtracting a linear fit. Bottom panel: the same with 
the parabolic fit.}
\label{dyhresfou}
\end{center}
\end{figure}

\noindent The parabolic fit has the form

$$O-C=-0.0001(2)+3.9(1)\times10^{-7}E-8.3(11)\times10^{-13}E^2$$

\noindent with an rms of 0\fd00115 ($\pm$errors of the last digits are in
parentheses). The linear fit yields higher rms (0\fd00157) and
the Fourier spectrum of the residuals (Fig.\ \ref{dyhresfou}) shows that 
the parabolic residuals are closer to an
uncorrelated noise. The peak in the spectrum of the linear residuals appears close
to the pulsational frequency (with wich the O$-$C was calculated): the difference 
is $2.30\times10^{-4}$ d$^{-1}$. Therefore, the hypothetical binary nature
(P\'ocs \& Szeidl 2000) of the star is not confirmed and we are confident 
about the definite detection of the slow period decrease. 
The parabolic coefficient (usually denoted as $\beta/2$=0.5~P~dP/dt) 
results in a period change ${1 \over P}{dP \over dt}=(-2.8\pm0.4)\times
10^{-8}$ year$^{-1}$, which is slightly smaller and more accurate than the 
value used by BP98 ($-4\times10^{-8}$ year$^{-1}$). Nevertheless, the main
conclusion drawn by several earlier authors is not changed, as this smaller
decreasing rate is generally in good agreement with the evolutionary 
calculations.

\subsection{YZ~Bootis}

YZ~Boo has quite a long observational record and the most relevant information
on the star was summarized by Pe\~na et al. (1999). Its period change was studied
by Szeidl \& Mahdy (1981), Jiang (1985) and Hamdy et al. (1986). The latter
authors analysed the longest dataset available at that time and concluded 
that the O$-$C diagram can be fitted with a positive parabola, indicating a 
constant period increase. They arrived at a relative period changing rate 
of ${1 \over P}{dP \over dt}=+3\times10^{-8}$ year$^{-1}$. 

\begin{table}
\begin{center}
\caption{Journal of observations and new times of maximum for YZ~Boo}
\label{yzobs}
\begin{tabular}{|lllrl|}
\hline
Date       & filter & Inst. & Points & $HJD_{\rm max}$\\
\hline
2001-03-16 & $V$  & Sz40  & 279 & 2451985.6116\\
2001-04-01 & $V$  & Sz40  & 536 & 2452001.5381\\
2001-04-25 & $V$  & Sz40 & 177 & 2452025.3695\\
2001-04-28 & $V$  & Sz40  & 697 & 2452028.3902\\
           &      &       &     & 2452028.4946\\
	   &      &       &     & 2452028.5975\\
2001-04-30 & $V$  & Sz40 & 526 & 2452030.3693\\ 
           &      &      &     & 2452030.4726\\
	   &       &     &     & 2452030.5759\\
2001-05-09 & $V$  & Sz40 & 307 & 2452039.3221\\
\hline
\end{tabular}
\end{center}
\end{table}

\begin{figure}
\begin{center}
\leavevmode
\psfig{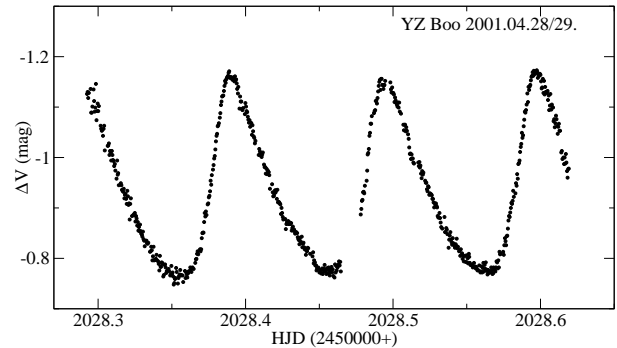}
\caption{Sample light curve for YZ~Boo.}
\label{yzlc}
\end{center}
\end{figure}

Our observations were carried out using Sz40 on six nights in 2001 
(Table\ \ref{yzobs}). The exposure time was between 15 and 40 s. Unfortunately,
only one comparison star was available in the small FOV of the instrument
(GSC 2569-1184 with $V\approx$11\fm3). Slightly more than 2500 data points were
obtained. The longest subset, containing 697 points, is shown 
in Fig.\ \ref{yzlc}. 

\begin{figure}
\begin{center}
\leavevmode
\psfig{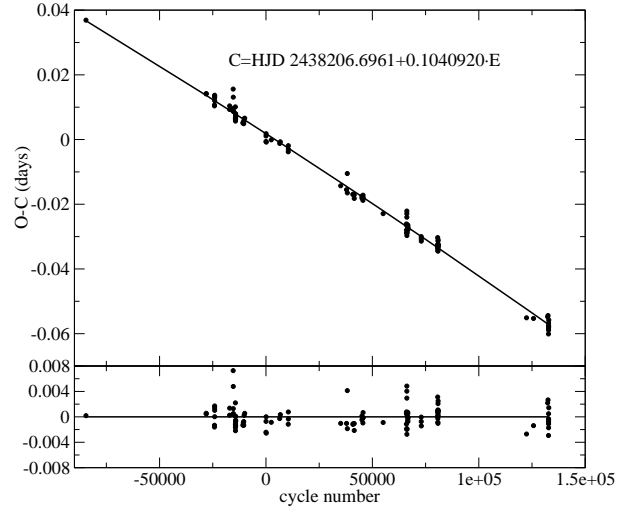}
\caption{The O$-$C diagram of YZ~Boo. The solid line denotes the 
parabolic fit. Residuals are shown below.}
\label{yzoc}
\end{center}
\end{figure}

\begin{figure}
\begin{center}
\leavevmode
\psfig{figure=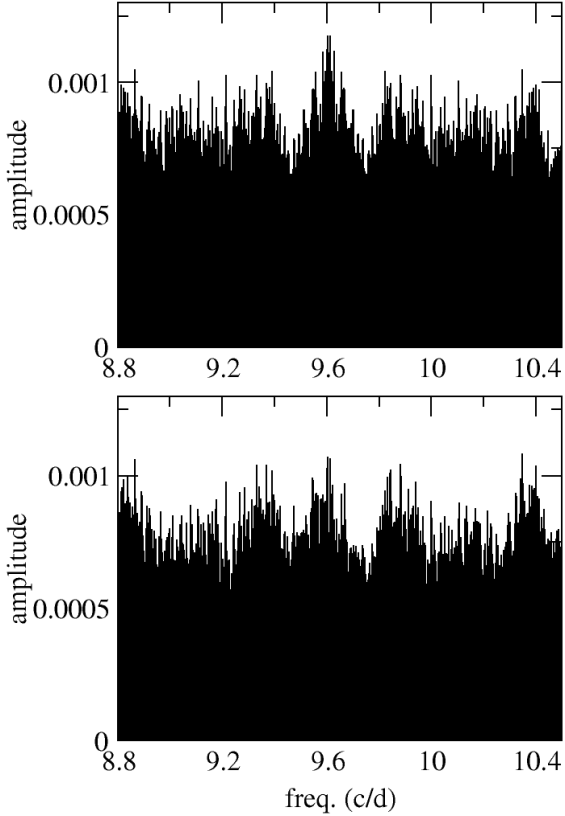,width=8cm}
\caption{Top panel: Fourier spectrum of the residuals for YZ~Boo
after subtracting a linear fit. Bottom panel: the same with 
the parabolic fit.}
\label{yzresfou}
\end{center}
\end{figure}

Ten new times of maximum were determined from the individual light curves
(Table\ \ref{yzobs}), which extended the existing data by another $\sim$15
years. Published times of maximum were collected from Hamdy et al. (1986) and 
references therein, Kim \& Joner (1994), Agerer et al. (1999) and 
Agerer \& H\"ubscher (2000). The O$-$C diagram  
was calculated with the following ephemeris: 

$$HJD_{\rm max}=2438206.6961+0.1040920\times E$$

\noindent and is plotted in Fig.\ \ref{yzoc}. The solid line corresponds to the
parabolic fit:

$$O-C=0.0018(2)-4.25(6)\times10^{-7}E-1.5(6)\times10^{-13}E^2$$

\noindent with an rms of 0\fd00169. The data can be fitted almost equally
well with a simple linear term (rms=0\fd00173), but a frequency analysis of the 
residuals slightly favours continuous period change. It is shown in
Fig.\ \ref{yzresfou}. There is a peak at $f\approx9.6116$
d$^{-1}(\approx{1 \over P}+4.7\times10^{-3}$ d$^{-1}$)
in the spectrum of the residuals of the linear fit, while
the spectrum of the parabolic residuals are more noise-like. This suggest 
a real period change over the six decades of observations, although 
the evidence is fairly weak.

If we accept its significance, the given second-order coefficient 
corresponds to a relative period changing rate 
${1 \over P}{dP \over dt}=(-1\pm0.4)\times10^{-8}$ year$^{-1}$, which has an 
absolute value of about one third of the previous estimate by Hamdy et al. 
(1986) and slightly larger than that of Peniche et al. (1985), but
with the opposite sign. It means that contrary to Hamdy et al. (1986) and 
Peniche et al. (1985), we found period decrease instead of increase. The
determined rate has great uncertainty and we recall a
note by Peniche et al. (1985): ``...it is impossible right now to decide 
if the period is constant or if it is varying, but this will be feasible only if
the star is regularly observed during the next 40 years''. Almost 
20 years have passed since then and the uncertainty is still too large. That 
clearly shows the need for continuous follow-up in the future.
The given slow rate is in good agreement with theoretical calculations
of BP98, which predict small period changes of the same order of magnitude.

\subsection{XX~Cygni}

Our unfiltered observations were published in Kiss \& Derekas (2000). 
Recently, XX~Cygni has been revisited by Zhou et al. (2002) and Blake et al. 
(2003). The history and observational record 
of XX~Cyg can be found in those papers. Contrary to our earlier conclusion, 
these authors provided convincing arguments for slow continuous period 
increase with at a rate of about ${1 \over P}{dP \over dt}=+1.13\times10^{-8}$
year$^{-1}$ (Blake et al. 2003) and $0.94\times10^{-8}$ year$^{-1}$ 
(Zhou et al. 2002). Here we add one night of observations carried out in 2002.
The comparison stars were the same as in Kiss \& Derekas (2000).

\begin{table}
\begin{center}
\caption{Journal of observations and new times of maximum for XX~Cyg}
\label{xxobs}
\begin{tabular}{|lllrl|}
\hline
Date       & filter & Inst. & Points & $HJD_{\rm max}$\\
\hline
2002-07-20 & $V$    & Sz40  & 60 & 2452476.4983\\
2002-07-20 & $R_{\rm C}$ & Sz40 & 61 & 2452476.4982\\
2002-07-20 & $I_{\rm C}$ & Sz40 & 59 & 2452476.4976\\
\hline
\end{tabular}
\end{center}
\end{table}

\begin{figure}
\begin{center}
\leavevmode
\psfig{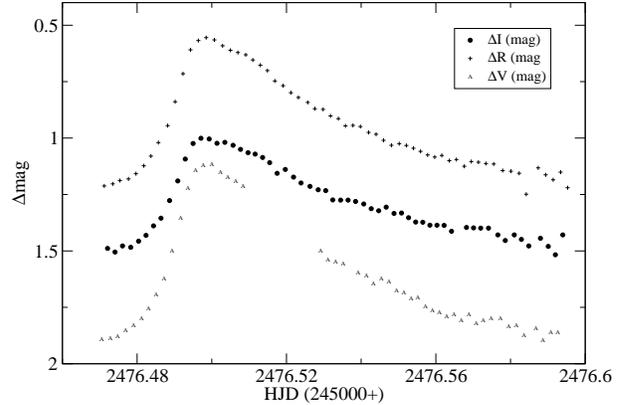}
\caption{$V(RI)_{\rm C}$ light curves of XX~Cyg.}
\label{xxlc}
\end{center}
\end{figure}

The obtained light curves are plotted in Fig.\ \ref{xxlc}, while the determined 
epochs of maximum are shown in Table\ \ref{xxobs}. The mean O$-$C value with the 
ephemeris from Szeidl \& Mahdy (1981) is $+0\fd0079$ which falls close to any
recent prediction. The light curve shape is similar to the ones shown
in Blake et al. (2003) and a well-expressed post-maximum bump is present in the 
$R$-band data. Only weak evidence is present for the possible transient
secondary maximum that is discussed in Blake et al. (2003): there is a slight 
hump around HJD 2452476.54 which resembles that of in Fig.\ 2 of Blake et al.
(2003) but no conclusive remark can be given based on this.

\subsection{SZ~Lyncis}

SZ~Lyncis is the best documented example of a binary pulsating star with very
clear light-time effects. The period variations of the star were studied by
Papar\'o et al. (1988) and Moffett et al. (1988). Since then it has been totally
neglected and no new times of maximum have appeared in the
literature, except the one by the Hipparcos satellite.

\begin{figure}
\begin{center}
\leavevmode
\psfig{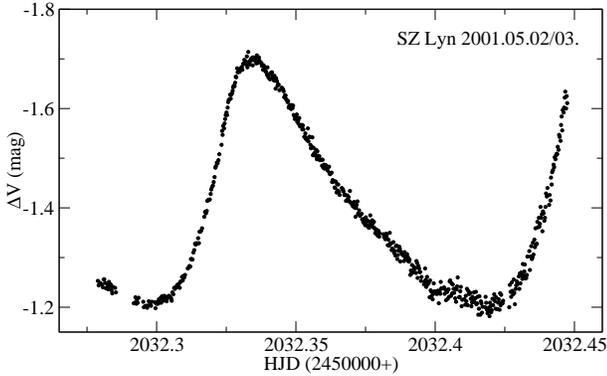}
\caption{Sample light curve for SZ~Lyn.}
\label{szlc}
\end{center}
\end{figure}

We observed SZ~Lyn with two telescopes on four nights in 2001. The exposure times
were between 15 and 120 seconds, depending on the instrument and weather
conditions. Two nearby stars were chosen as comparisons (comp = GSC 2979-1329,
$V\approx10\fm5$, check = GSC 2979-1343, $V\approx10\fm8$). More than 1500
individual points have been acquired (Table\ \ref{szobs}) and a sample light 
curve is shown in Fig.\ \ref{szlc}.

\begin{table}
\begin{center}
\caption{Journal of observations and new times of maximum for SZ~Lyn}
\label{szobs}
\begin{tabular}{|lllrl|}
\hline
Date       & filter & Inst. & Points & $HJD_{\rm max}$\\
\hline
2001-02-25 & $V$    & P60   & 395    & 2451966.5244\\
2001-02-26 & $V$    & P60   & 239    & 2451967.3683\\
2001-04-03 & $V$    & Sz40  & 287    & 2452003.4072\\
2001-05-02 & $V$    & Sz40  & 619    & 2452032.3351\\
\hline
\end{tabular}
\end{center}
\end{table}

\begin{figure}
\begin{center}
\leavevmode
\psfig{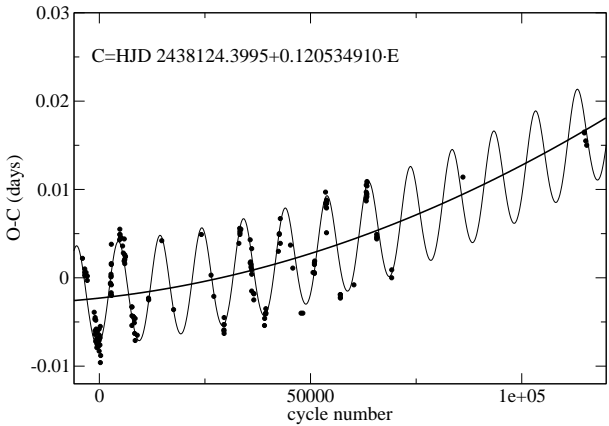}
\caption{The O$-$C diagram of SZ~Lyn. The thick solid line shows the 
parabolic part of the fit.}
\label{szoc}
\end{center}
\end{figure}

The four new times of maximum supplemented with the Hipparcos epoch
(HJD 2448500.0560, ESA 1997) were added to the list of maxima
(Papar\'o et al. 1988, Moffett et al. 1988). The resulting O$-$C diagram is 
plotted in Fig.\ \ref{szoc}. It was calculated with the ephemeris

$$HJD_{\rm max}=2438124.39955+0.120534910 \times E$$

\noindent taken from Papar\'o et al. (1988). 
Our new observations confirm the consistent picture of the 
light-time effect thoroughly discussed in Papar\'o et al. (1988)
and Moffett et al. (1988). The parabolic and light-time fit resulted in
essentially the same parameters as those of the mentioned papers, which is 
not surprising, as the overwhelming majority of the fitted data is the same.
We could, however, improve the value of the orbital period, which
turned out to be $P_{\rm orb}=1190\pm5$ d (contrary to 1177\fd7 and
1181\fd1 by Papar\'o et al. 1988 and Moffett et al. 1988, respectively).
On the other hand, 
we derived a slightly smaller period changing rate of
${1 \over P}{dP \over dt}=(+5\pm1.8)\times10^{-8}$ year$^{-1}$. We note 
that the quadratic term is still too uncertain. We have experimented with 
different averaging of the original O$-$C points to decrease the effects 
of the unequal data distribution. Different binnings changed a lot
the parabolic coefficient (almost by a factor of two), and that marks
a certain limit in the data interpretation.

\subsection{BE~Lyncis}

Kiss \& Szatm\'ary (1995) suspected binarity of BE~Lyn, based 
on the seemingly cyclic O$-$C diagram. In order to confirm this 
preliminary result, we have continuously monitored this star since
then. Previous observations are summarized in Rodr\'\i guez (1999), who did 
not find any long-term amplitude change of the light curve.

\begin{table*}
\begin{center}
\caption{Journal of observations and new times of maximum for BE~Lyn.
The first five epochs were re-determined from the observations in
Kiss \& Szatm\'ary (1995). For multicolour observations we only list
epochs based on $V$-band data (either from Johnson or Str\"omgren photometry).}
\label{beobs}
\begin{tabular}{|lllrl|lllrl|}
\hline
Date       & filter & Inst. & Points & $HJD_{\rm max}$ &  Date       & filter & Inst. & Points & $HJD_{\rm max}$\\
\hline
1995-01-31 & $BV$   & Sz40  & 267 &  2449749.4651& 1999-03-08 & $BV$  & Sz40  & 172 &  2451246.2747\\
           &        &       &     &  2449749.5622&	     &        &       &     &  2451246.3704\\
	   &        &       &     &  2449749.6562&	     &        &       &     &  2451246.4674\\
1995-02-05 & $UBV$  & Sz40  & 48  &  2449754.3545& 1999-03-08 & $uvby$ & SNO90 & 62  &  2451246.3702\\
1995-02-13 & $UBV$  & Sz40  & 63  &  2449762.4075&	     &        &       &     &  2451246.4675\\
1996-01-31 & $BV$   & Sz40  & 51  &  2450114.5363& 1999-04-24 & $uvby$ & SNO90 & 60  &  2451293.3477\\
1996-02-25 & $BV$   & Sz40  & 60  &  2450139.4639&	     &        &       &     &  2451293.4437\\
1996-02-26 & $BV$   & Sz40  & 60  &  2450140.4208& 1999-04-25 & $uvby$ & SNO90 & 18  &  --\\
1997-02-20 & $BV$   & Sz40  & 75  &  2450500.5075& 1999-05-12 & $uvby$ & SNO90 & 56  &  2451311.4673\\
1998-01-30 & $BV$   & Sz40  & 114 &  2450844.3906& 1999-05-14 & $uvby$ & SNO90 & 37  &  2451313.4802\\
           &        &       &     &  2450844.4860& 2000-02-10 & $BV$  & Sz40  & 57  &  2451585.3663\\
1999-01-26 & $uvby$ & SNO90 & 99  &  2451205.5313& 2001-01-19 & $uvby$ & SNO90 & 33  &  2451929.4424\\
           &        &       &     &  2451205.6277&	     &        &       &     &  2451929.5377\\
	   &        &       &     &  2451205.7231& 2001-02-24 & $B$   & P60   & 215 &  2451965.2974\\
1999-02-27 & $BV$   & Sz40  & 234 &  2451237.2625& 2001-02-24 & $V$   & P60   & 830 &  2451965.3927\\
           &        &       &     &  2451237.3596&	     &        &       &     &  2451965.4896\\
	   &        &       &     &  2451237.4552& 2001-03-14 & $V$   & Sz40  & 224 &  --\\
	   &        &       &     &  2451237.5500& 2001-03-16 & $V$   & Sz40  & 438 &  2451985.3329\\
1999-03-01 & $BV$   & Sz40  & 149 &  2451239.3712&	     &        &       &     &  2451985.5235\\
           &        &       &     &  2451239.4684& 2001-12-08 & $V$   & Sz40  & 193 &  2452252.5191\\
1999-03-02 & $uvby$ & SNO90 & 305 &  2451240.4268& 2002-04-06 & $V$   & Sz28  & 182 &  2452371.4005\\
           &        &       &     &  2451240.5242&	     &        &       &     &  2452371.4946\\
	   &        &       &     &  2451240.6200& 2002-05-01 & $V$   & Sz40  & 1012 & 2452396.3266\\
1999-03-05 & $BV$   & Sz40  & 128 &  2451243.3020&	     &        &       &      & 2452396.4225\\
           &        &       &     &  2451243.4008&	      &        &       &      & 	    \\
\hline
\end{tabular}
\end{center}
\end{table*}

All four instruments of this project have been used for photoelectric and CCD
observations of BE~Lyn. The full journal of observations and the list of 
epochs of maximum are given in Table\ \ref{beobs}. Because of the different
detectors, various comparison stars were chosen. For photoelectric photometry 
with Sz40, HIP 45515 ($V=9\fm30$, spectral type F8),  for CCD photometry with
Sz40 and P60, GSC 3425-0544 ($V\approx11\fm1$) were used. For photoelectric
photometry with SNO90, HD~79763 ($V=5\fm95$) served as the comparison
and HD~80079 ($V=6\fm90$) and HD~79439 ($V=4\fm82$) as the check stars. 
Integration times varied between 8 and 120 seconds, depending on the instrument,
detector and weather conditions. 

\begin{figure}
\begin{center}
\leavevmode
\psfig{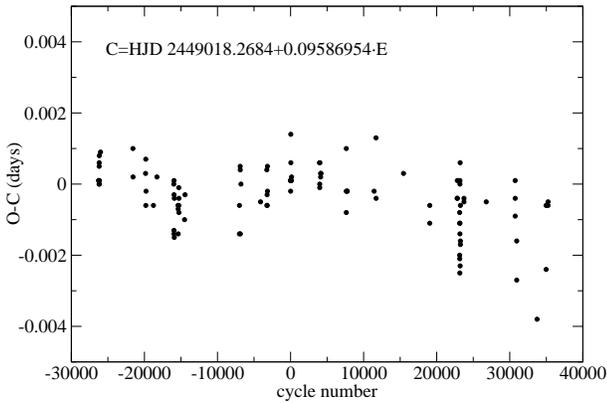}
\caption{The O$-$C diagram of BE~Lyn}
\label{beoc}
\end{center}
\end{figure}

The 47 new epochs supplemented the earlier O$-$C diagram, and the final  dataset
consists of 106 points (references can be found in Kiss \& Szatm\'ary 1995). 
We plot the resulting O$-$C diagram in Fig.\ \ref{beoc}. It has been calculated
with the following ephemeris:

$$HJD_{\rm max}=2449018.2684+0.09586954\times E$$

\noindent (Liu \& Jiang 1994). Obviously, the LITE-solution of Kiss \& Szatm\'ary
(1995) does not hold, as there is no cyclic feature in the present diagram. Slight
variations are suggested but neither a linear nor a parabolic fit describes 
very well the general appearance. 

\begin{figure} 
\begin{center} 
\leavevmode 
\psfig{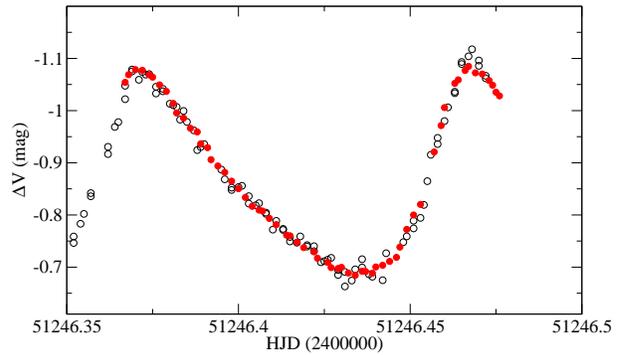}
\caption{A comparison of two simultaneous datasets of BE~Lyn obtained with Sz40
(open circles) and SNO90 (filled circles).} 
\label{sz_sno} 
\end{center}
\end{figure}

We have made several tests to improve our understanding of the limitations in 
the case of BE~Lyn. First, we have checked the consistency of the determined epochs
of  maxima with simultaneously obtained observations. On March 8, 1999, there
were two cycles observed in parallel with Sz40 and SNO90. We show the light curve
comparison in Fig.\ \ref{sz_sno}. Spanish data were shifted vertically to the
best match with the Hungarian observations. The independent measurements yielded
almost the same times of maximum, both were within 0\fd0002 (see Table\
\ref{beobs}). So we believe that the estimated accuracy of $\pm$0\fd0003 for the 
individual epochs is indeed a realistic value. There is, however, a larger
scatter in the O$-$C diagram which appears to be larger than the observational 
noise. Fourier analysis of the diagram did not help, because both the residuals
of the linear fit and the parabolic fit (see below) showed peaks close to the
pulsational frequency, implying higher order period change (that is a simple
consequence of the curved nature of the diagram) or low-frequency components in
the spectrum caused by the inappropriate fit. 

\begin{figure*}
\begin{center}
\leavevmode
\psfig{figure=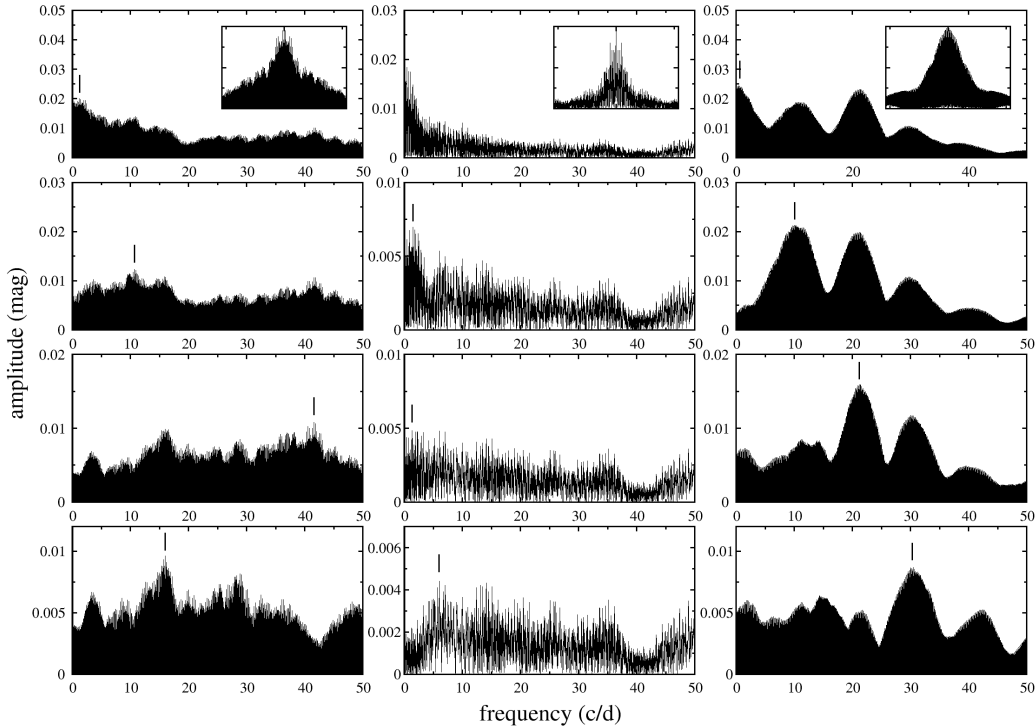,width=14cm}
\caption{Subsequent Fourier spectra of three subsets (1995-1997, 1999 and 2001)
of BE~Lyn (three columns) after prewhitening with the dominant frequency and its
harmonics. The highest peaks are marked in every spectrum, with which the next 
prewhitening steps
were done, from top to bottom. The small inserts show the window functions. 
Although there is some evidence for weak secondary periodicity, no 
features can be identified as being the same in all three subsets. 
The S/N ratios (Breger et al. 1993) of the marked peaks range from 3 to 10.}
\label{befou}
\end{center}
\end{figure*}

Secondly, we have searched low-amplitude secondary periodicity in the light
curves. Garrido \& Rodr\'\i guez (1996) have already tried to detect
microvariability of BE~Lyn but found nothing, based on five nights of
observations obtained with a span of two years. 
Since we have 27 $V$-band light curves
distributed almost uniformly from 1995 to 2002, we performed their Fourier
analysis with the same aim. We have divided the data into three
subsets, in which the changes of the dominant frequency can be neglected. Nightly
mean values have been subtracted (as did, e.g., Kiss et al. 2002a) and that is
why we had to exclude a few nights with short light curve coverage from the
analysis. The first subset covered 1995-01-31--1997-02-20; the second 
1999-01-26--1999-05-14; and the third 2001-02-24--2002-05-01. The most 
homogeneous and best-quality subset is the second one, which consists mainly 
of high-precision photoelectric photometry with SNO90. 

We prewhitened each subset with $f_0=10.430842$ d$^{-1}$ and its
harmonics up to the fifth order and analysed the residuals for periodicity. The
results can be summarized as follows. In every subset, we detected low-frequency
($f<2$ d$^{-1}$) components which we attributed to nightly zero-point
uncertainties. After their removal, we could not infer any periodicity that
was common to all subsets. The prewhitened spectra in Fig.\ \ref{befou} show that,
although there are some hints of multiperiodicity, no identical features can be 
identified in all three subsets. Most importantly, the second subset 
(middle column in Fig.\ \ref{befou}) does not reveal anything with amplitude 
larger then 5 mmag. Therefore, we conclude that: {\it i)}
there is no detectable stable secondary periodicity in the light curve of BE~Lyn
above the mmag level; {\it ii)} if the scatter of the O$-$C diagram is real than
it implies some light curve instabilities (transient phenomena) which
are not strictly periodic but affect the light curve shape. Nevertheless, we
admit that the whole effect is just on the limit of detectability and no
unambiguous conclusion can be drawn at the present time. And, as noted by Garrido
\& Rodr\'\i guez (1996), the microvariability can be studied only with the
highest quality observations (long-term homogeneous photometry with internal
precision of a few mmag), which is, unfortunately, not the case for our data. 

We have also checked the Fourier amplitude and phase parameters, used to 
describe the light curve shape, for seasonal variability. We have checked 
$R_{21}$, $R_{31}$, $\varphi_{21}$ and $\varphi_{31}$ (see, e.g., Poretti 2001)
but found no changes. For instance, $R_{21}$ was 0.335, 0.328 and 0.303, 
while $R_{31}$ varied from 0.108, 0.111, 0.113. As a comparison, the uncertainty
of these values is a few in the last digit and we do not consider the mentioned
variations as significant. Thus we exclude the presence of long-term 
light-curve shape changes. 

By setting aside the higher-order period change, we have determined an ephemeris 
correction with a linear fit and estimated the period changing
rate with a parabolic fit. The linear approximation yields the following 
improved ephemeris:

$$HJD_{\rm max}=2449018.2681+0.095869521(4)\times E$$

\noindent which can be used to predict forthcoming maxima. The parabolic fit
gives  $\beta=(-1.3\pm0.5)\times10^{-12}$, that would correspond to 
${1 \over P}{dP \over dt}=(-5\pm1.9)\times10^{-8}$ year$^{-1}$. However, we 
consider this value only as an upper limit for the period change, as the present
situation is not as clear as for the previous examples. Nevertheless, this 
limit is in good agreement with theoretical expectations of BP98.

\subsection{DY~Pegasi}

DY~Peg is one of the best observed HADSs, with observations dating back 
more than a half century. A recent update of the period change was published 
by Blake et al. (2000), while the physical parameters of the star were 
determined by Wilson et al. (1998) and Pe\~na et al. (1999). Garrido \&
Rodr\'\i guez (1996) presented some evidence for secondary periodicity,
although based on only four nights of observations. The most detailed 
period study is that of Koen (1996), revealing the complex nature of the 
period variations. Interestingly, even the seemingly simple parabolic O$-$C 
descriptions yielded ambiguous results: Blake et al. (2000) estimated a 
quadratic term of the O$-$C diagram almost a factor of seven less than, e.g.
Mahdy (1987) ($-1.36\times10^{-12}$ vs. $-0.231\times10^{-12}$). 

\begin{table}
\begin{center}
\caption{Journal of observations and new times of maximum for DY~Peg}
\label{dypobs}
\begin{tabular}{|lllrl|}
\hline
Date       & filter & Inst. & Points & $HJD_{\rm max}$\\
\hline
2001-01-01 & $B$    & P60  & 101 & 2451911.2338\\
           &        &      &     & 2451911.3062\\
2001-08-03 & $V$    & Sz40 & 166 & 2452125.5619\\
2001-08-04 & $V$    & Sz40 & 152 & 2452126.5833\\
2001-08-12 & $V$    & Sz40 & 253 & 2452134.5332\\
           &        &      &     & 2452134.6062\\
2001-08-15 & $V$    & Sz40 & 339 & 2452137.5228\\
           &        &      &     & 2452137.5955\\
2001-09-05 & $uvby$ & SNO90 & 57 & 2452158.3805\\
2001-09-08 & $uvby$ & SNO90 & 69 & 2452161.4431\\
\hline
\end{tabular}
\end{center}
\end{table}

\begin{figure}
\begin{center}
\leavevmode
\psfig{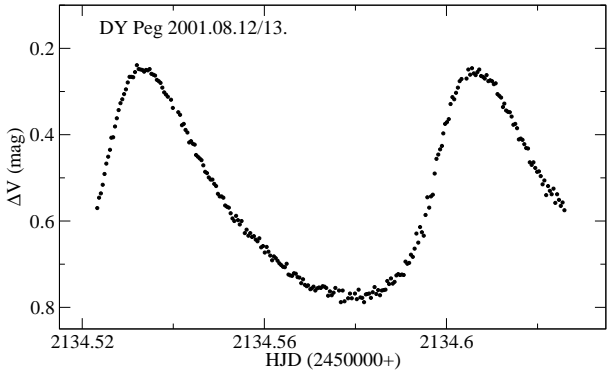}
\caption{Sample light curve for DY~Peg.}
\label{dyplc}
\end{center}
\end{figure}

Our observations were carried out at three observatories on seven nights in
2001. The full log of observations, supplemented with the 
new times of maximum, is presented in Table\ \ref{dypobs}. Throughout the 
observations we used the same comparison star with all of the instruments
(HD~218587, $V=9\fm80$, $b-y=0\fm376$, $m_1=0\fm180$, $c_1=0\fm399$,
as given in the SIMBAD database). The secondary comparison was 
either GSC 1712-0542 ($V\approx11\fm7$) or GSC 1712-1246 ($V\approx11\fm1$).
The integration time varied between 30 and 120 seconds. A sample light curve is 
shown in Fig.\ \ref{dyplc}.

\begin{figure}
\begin{center}
\leavevmode
\psfig{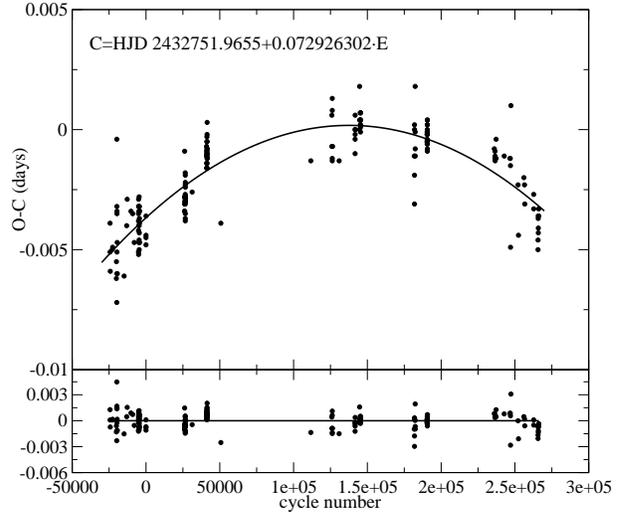}
\caption{The O$-$C diagram of DY~Peg. The solid line denotes the 
parabolic fit.}
\label{dypoc}
\end{center}
\end{figure}

After collecting and supplementing the full list of maxima (Blake et al.
2000 and references therein, Van Cauteren \& Wils 2000, Agerer et al. 2001)
we show in Fig.\ \ref{dypoc} the final O$-$C diagram covering 1943--2001 with
the ephemeris of Mahdy (1987):

$$HJD_{\rm max}=2432751.9655+0.072926302 \times E$$

\noindent The parabolic fit has to following form:

$$O-C=-0.0037(1)+5.6(2)\times10^{-8}E-2.0(1)\times10^{-13}E^2$$

\noindent with an rms of 0\fd001. This results in 
${1 \over P}{dP \over dt}=(-2.79\pm0.14)\times 10^{-8}$ year$^{-1}$,
which is in good agreement with Mahdy (1987) and Pe\~na et al.
(1987), thus we do not confirm the different result of Blake et al. (2000).
We also note that the updated O$-$C diagram is obviously not fully
described by the parabolic fit; for instance, the last $\sim25000$ cycles
would need a much steeper function than the present parabola. Therefore, 
we fully agree with Koen (1996) that viable quantitative alternatives
to the traditional O$-$C analysis must be used when investigating the 
fine details of the period changes. Here we did not go beyond the 
simplest analysis, because that would lead us too far from the 
primary purposes of this paper.

\subsection{GW~Ursae Majoris}

The light variations of GW~UMa ($\langle V\rangle\approx 9.7$,
$A_{\rm V}=0\fm44$, $P=0\fd2032$) were discovered by the Hipparcos 
satellite, and despite its brightness and short period,
there has been no new observation since the discovery. A special 
interest can be attributed to this star due to the period value:
it is just on the boundary between HADSs and the shortest period
RR~Lyrae stars (Poretti 2001). In order to place the star in the light
curve diagnostic diagrams used by Poretti (2001), we observed the star
in $B$, $V$ and $I_{\rm C}$ bands on 6 nights in 2002 (Poretti utilized
I-band data of hundreds of stars to infer characteristic Fourier-parameters
of HADS and RRc variables). More than 2500 individual points were obtained
(the full observing log is presented in Table\ \ref{gwobs}). The comparison
star was GSC 3011-2535 ($V\approx10\fm9$).

\begin{table}
\begin{center}
\caption{Journal of observations and new times of maximum for GW~UMa}
\label{gwobs}
\begin{tabular}{|lllll|}
\hline
Date       & filter & Inst. & Points & $HJD_{\rm max}$\\
\hline
2002-04-30 & $V$    & Sz40  & 820 & 2452395.5408\\
2002-05-01 & $V$    & Sz40  & 180 & 2452396.5584\\
2002-05-02 & $I_{\rm C}$ & Sz40 & 674 & 2452397.3702\\
           &             &      &     & 2452397.5782\\
2002-05-03 & $V$    & Sz40  & 229 & 2452398.3850:\\
2002-05-03 & $I_{\rm C}$ & Sz40 & 258 & 2452398.3877\\ 
2002-05-22 & $B$    & Sz40 & 181 & 2452417.4858\\
2002-05-31 & $B$     & Sz40 & 204 & 2452426.4259\\
\hline
\end{tabular}
\end{center}
\end{table}

\begin{figure}
\begin{center}
\leavevmode
\psfig{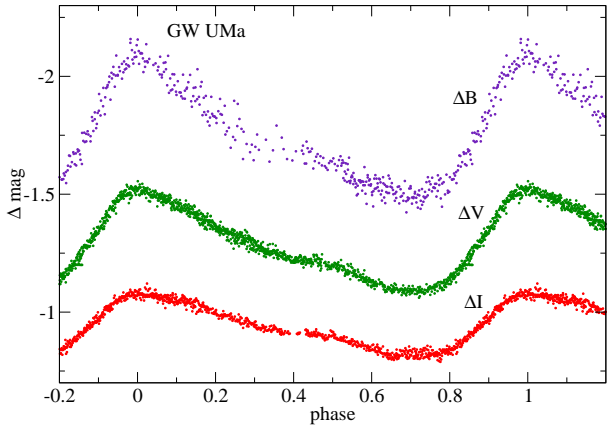}
\caption{The phase diagrams of GW~UMa ($E_0$=2452395.5408, $P=0\fd2031940$).}
\label{gwfaz}
\end{center}
\end{figure}

Our analysis showed that the star is likely to be monoperiodic, as we could not
detect cycle-to-cycle changes of the light curve shape in any band (although this
conclusion is based on a fairly scanty dataset). The phased light curves are
shown in Fig.\ \ref{gwfaz}. The amplitudes in the instrumental $BVI_{\rm C}$ 
system are 0\fm62, 0\fm44 and 0\fm27, respectively (all but one night of 
observations were single-filtered, for better time resolution, but the trade-off
is the lack of standard transformations). The Fourier amplitude 
parameter $R_{21}$=0.40 for the
$I$-band data placed the star among the HADSs (see Fig.\ 7 in Poretti 2001).
Therefore, we exclude the possibility of first- or second-overtone RR~Lyrae
pulsation, as suggested  for other short-period pulsators of similar light curve
shapes and periods  (see, e.g. Kiss et al. 1999 for a discussion of 
the nature of V2109~Cyg which, however, recently turned to be a multiperiodic HADS 
(Rodr\'\i guez, in preparation)). Eight new times of maximum were
determined (Table\ \ref{gwobs}) and the  mean O$-$C value with the  Hipparcos
ephemeris ($E_0=2448500.1160$, $P=0\fd2031940$) is $-0\fd0062$  (c.c. $-$9 min),
suggesting either inaccurate Hipparcos period, or slight period change during the
last decade. Assuming a constant period, we calculated to following corrected
ephemeris:

$$HJD_{\rm max}=2448500.1160+0.20319367(3)\times E$$

\noindent In any case, further CCD observations are needed to clarify the
situation. The star is probably a Pop. II object, as the high galactic latitude 
($+59\fdg14$) is associated with large radial velocity  (the SIMBAD database
lists $-81$ km~s$^{-1}$). Unfortunately, the Hipparcos parallax is not useful
($0.58\pm2.88$ mas), thus no other information can be deduced from the 
presently available data.

\section{Summary}

In this paper, we presented observational results of an almost 8-year long
project, with the main goal of updating our knowledge on the period change
of selected bright northern HADSs. Our results strengthen 
the generally adopted view of evolutionary period changes, and the determined 
period decreasing and increasing rates are in good accordance with 
the theory. The most relevant new results are mostly negative results,
in the sense of disproving earlier conclusions that appeared in the literature:

\begin{enumerate}

\item For DY~Herculis, we do not confirm the suspected 
binarity (P\'ocs \& Szeidl 2000). The updated O$-$C diagram between 1938 and 2001
does not show clear signs of a light-time effect. Instead of that, a 
quadratic fit is fairly satisfactory. 

\item For YZ~Bootis, we do not confirm the increase of the period (Hamdy et al.
1986). Instead of that we find slow period decrease, close to the limits of 
detection.

\item For BE~Lyncis, we do not confirm the suspected 
binarity (Kiss \& Szatm\'ary 1995). The updated O$-$C diagram between 1986 
and 2002 suggests complex period variations with no real cyclic nature. 
Fourier-analysis of the light curves did not reveal possible multiple
periodicity above the millimag level. A formal quadratic fit to the
O$-$C diagram resulted in an upper limit to the period change rate that is
in agreement with the theoretical expectations. Further observations
of the star are needed. 

\item For DY~Peg, we do not confirm the results 
on its smooth period change rate (Blake et al. 2000). Although there are 
hints for more complex period variation (see also Koen 1996), the updated
O$-$C diagram between 1943 and 2001 can be very well fitted with a parabola that
has similar coefficients than in Mahdy (1987) or Pe\~na et al. (1987). 

\item We presented the first observations of GW~UMa since its
discovery by the Hipparcos satellite. Empirical evidence was found for 
its HADS nature and the available kinematic data suggest the star to be 
a Pop. II object (thus possibly belonging to SX~Phe stars). 
The period of 0\fd203 makes the classification an intriguing task, as 
GW~UMa is right on the border between RR~Lyrae and HADS stars.
The stability of the light curve shape excludes the possibility of a
relatively high-amplitude secondary period.

\end{enumerate}

For the remaining two stars (XX~Cyg and SZ~Lyn), our conclusions
are more or less confirmations of previously published results. As expected
from the longer time basis available for the target stars, the newly determined 
quadratic coefficients (and the corresponding period changes) have better 
defined values and our calculations yielded a slight or significant 
decrease 
of the absolute value of the period changes. With these corrections, 
we find better agreement with theory.

Finally, let us point out again that HADS stars are easy 
targets for small astronomical instruments. The open questions 
raised for our programme stars or those of for other bright HADSs make 
continuous photometric monitoring a useful project 
for small telescopes under light-polluted urban environment. 
Even well-equipped amateur astronomers or small colleges can make 
significant contribution to observational studies of high-amplitude $\delta$
Scuti stars.

\begin{acknowledgements}
This work has been supported by the MTA-CSIC Joint Project No. 15/1998, 
FKFP Grant 0010/2001, OTKA Grants \#T032258, \#T034615, \#F043203,
\#T042509,  
Pro Renovanda
Cultura Hungariae Student Science Foundation and the Australian 
Research Council. LLK wishes to thank kind hospitality of Mr. Tam\'as 
Zalezs\'ak
and his wife, Szilvia when staying in Brisbane, Australia, where the final
modifications of the paper have been figured out. Thanks are also due to
Dr. T. Bedding for a careful reading of the manuscript. 
The NASA ADS Abstract Service was used to access data and references. This 
research has made use of the SIMBAD database, operated at CDS-Strasbourg, France.
\end{acknowledgements}

\end{document}